\documentclass{article}

\usepackage{arxiv}

\usepackage[utf8]{inputenc} % allow utf-8 input
\usepackage[T1]{fontenc}    % use 8-bit T1 fonts
\usepackage[colorlinks=false]{hyperref}       % hyperlinks
\usepackage{url}            % simple URL typesetting
\usepackage{booktabs}       % professional-quality tables
\usepackage{amsfonts}       % blackboard math symbols
\usepackage{nicefrac}       % compact symbols for 1/2, etc.
\usepackage{microtype}      % microtypography
\usepackage{lipsum}
\usepackage{graphicx}

\graphicspath{ {./images/} }

\title{Knowing Ourselves Through Others: Reflecting with AI in Digital Human Debates}

\author{
 Ichiro Matsuda\thanks{Both authors contributed equally to the paper.} \\
  University of Tsukuba\\
  Tsukuba, Ibaraki, Japan\\
  \texttt{ichi6m@digitalnature.slis.tsukuba.ac.jp} \\
   \And
 Komichi Takezawa\footnotemark[1]\\
  University of Tsukuba\\
  Tsukuba, Ibaraki, Japan\\
  \texttt{komichi@digitalnature.slis.tsukuba.ac.jp} \\
   \And
 Katsuhito Muroi \\
  University of Tsukuba\\
  Tsukuba, Ibaraki, Japan\\
  \texttt{katsuhitomuroi@digitalnature.slis.tsukuba.ac.jp} \\
   \And
 Kensuke Katori\\
  University of Tsukuba\\
  Tsukuba, Ibaraki, Japan\\
  \texttt{kenkenissocool@digitalnature.slis.tsukuba.ac.jp} \\
   \And
 Ryosuke Hyakuta\\
  Miscrosoft Japan Co., Ltd.\\
  Tokyo, Japan\\
  \texttt{rhyakuta@microsoft.com} \\
   \And
 Jingjing li\\
 University of Tsukuba\\
  Tsukuba, Ibaraki, Japan\\
  \texttt{li@digitalnature.slis.tsukuba.ac.jp} \\
   \And
 Yoichi Ochiai\\
  University of Tsukuba\\
  Tsukuba, Ibaraki, Japan\\
  \texttt{wizard@slis.tsukuba.ac.jp} \\
}
\date{November, 17, 2025}

\begin{document}
\maketitle
\begin{abstract}
LLMs can act as an impartial other, drawing on vast knowledge, or as personalized self-reflecting user prompts. These personalized LLMs, or Digital Humans, occupy an intermediate position between self and other. This research explores the dynamic of self and other mediated by these Digital Humans. Using a Research Through Design approach, nine junior and senior high school students, working in teams, designed Digital Humans and had them debate. Each team built a unique Digital Human using prompt engineering and RAG, then observed their autonomous debates. Findings from generative AI literacy tests, interviews, and log analysis revealed that participants deepened their understanding of AI's capabilities. Furthermore, experiencing their own creations as others prompted a reflective attitude, enabling them to objectively view their own cognition and values. We propose "Reflecting with AI"—using AI to re-examine the self—as a new generative AI literacy, complementing the conventional understanding, applying, criticism and ethics.
\end{abstract}

%% 本文
\section{Introduction}

According to Robin Dunbar's \textit{Vocal Grooming} hypothesis, language evolved as a substitute for physical grooming among primates, enabling social cohesion in larger groups~\cite{dunbar_grooming_2002}. Yet language transcends mere social bonding. In the ancient Greek agora, citizens established the art of resolving conflicts through dialogue rather than violence~\cite{Sennett2016Concentrating}. This ``persuasion through logical discourse'' became the foundation of debate.
\begin{figure}
    \centering
    \includegraphics[width=0.8\textwidth]{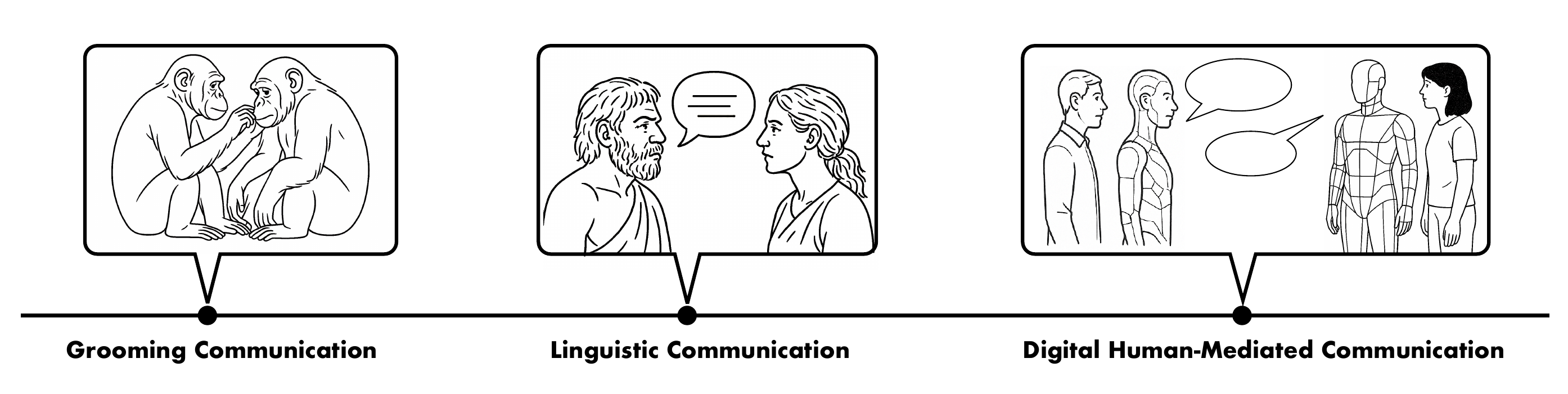}
    \caption{Vocal grooming is the origin of linguistic communication, which builds social relationships within groups. As LLMs can communicate with humans, we can expect them to build new relationships between humans and LLMs and create new social groups.}
    \label{fig:timeline}
\end{figure}

Debate has evolved beyond a mere framework for conflict resolution into competitive debate—an intellectual contest with rules, roles, and victory conditions. As Huizinga demonstrated in \textit{Homo Ludens}, play is a fundamental activity that creates culture~\cite{huizinga_homo_1949}, and debate exemplifies what Caillois categorized as both competition (agon) and mimicry~\cite{caillois_man_2001}. In competitive debate, participants adopt positions not necessarily aligned with their personal beliefs, anticipating opponents' logic while constructing refutations. Engaging in such structured debate cultivates critical thinking skills~\cite{ang_systematic_2019} and enables discovery through argumentation.

The emergence of generative AI, particularly Large Language Models (LLMs), has given humanity its first non-human entity capable of sophisticated language manipulation. This technological innovation introduces entirely new forms of participation in debate, previously exclusive to humans. In conventional debate, humans directly serve as argumentative agents. However, in the novel format we propose—Digital Human Debates (DHD)—the human role fundamentally transforms.

\begin{figure}[ht]
\centering
\includegraphics[width=0.8\textwidth]{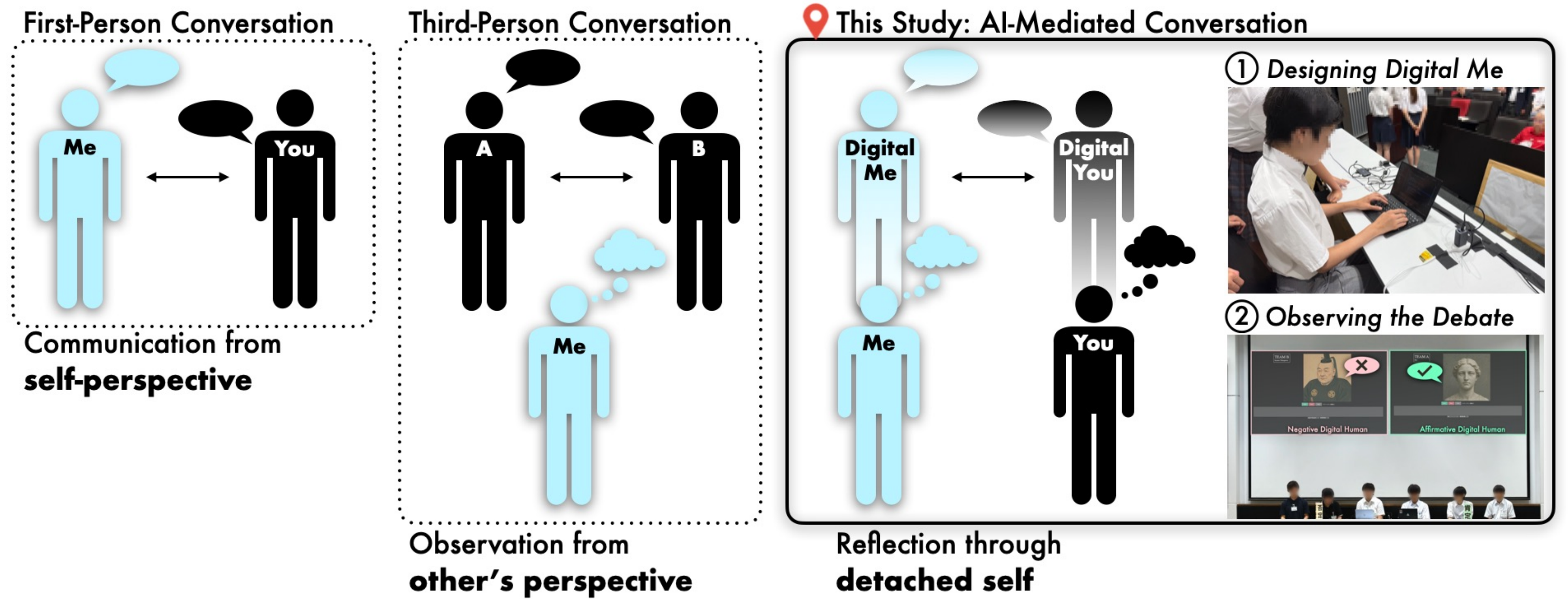}
\caption{Positioning of this research in conversation perspectives. \textit{First-Person Conversation}: Direct experiential engagement in communication with others. \textit{Third-Person Conversation}: Observing conversations from an external standpoint. \textit{This Study (Digital Human-Mediated Conversation)}: Designing digital humans that embody one's thinking patterns, then observing their dialogues from a third-person perspective—creating a unique hybrid of self-projection and external observation.}
\label{fig:conversation}
\end{figure}

In DHD, humans first become designers who embed their thinking patterns and logical structures into digital humans through prompt engineering. This is not merely configuration but creative play—crafting a \textit{semi-autonomous self} imbued with one's intentions. Once debate begins, humans shift from direct arguers to observers watching their designed digital humans engage in discourse. Crucially, while digital humans originate from designers' intentions, they generate unexpected logical developments in real-time. Participants experience these debates—conducted by entities that \textit{reflect their thinking yet are not themselves}—from a perspective that is neither pure observation of others nor complete self-projection. The impact of such self-projected digital human debates on the original designers remains unexplored.

Our research question investigates what cognitive and emotional experiences emerge when participants observe debates between digital humans they have designed their own intentions. Employing a Research Through Design approach, we conducted a study where nine junior and senior high school students designed digital humans in teams and observed autonomous debates between them. Through prompt engineering and knowledge input via RAG (Retrieval Augmented Generation), each team created digital humans with unique personalities and knowledge bases. Analysis of post-debate comprehension tests, interviews, and logs revealed that participants developed nascent reflective attitudes by experiencing their designed digital humans as others, enabling objective examination of their own cognition and values. This suggests possibilities beyond conventional AI literacy's understanding, application, and critique—a new form of AI literacy that reframes self-understanding through AI. We propose Reflecting with AI as a novel competency in generative AI literacy.

Our contributions are as follows:
\begin{itemize}
\item Implementation of \textit{Digital Human Debates} where participants design digital humans and observe their autonomous debates
\item Identification of \textit{reflection} as an emergent AI literacy competency through generative AI engagement
\end{itemize}

\section{Related Work}

\subsection{AI and Debate}

The earliest conversational agents employed rule-based methods, exemplified by Weizenbaum's ELIZA, which relied on simple pattern matching ~\cite{weizenbaum_elizacomputer_1966}. Subsequently, conversational agents found applications in gaming contexts, with systems like Seaman ~\cite{segawiki_seaman_2025} and Façade ~\cite{playabl_contact1_2025} emerging. To enable free-form dialogue, machine learning approaches including deep learning were introduced ~\cite{vinyals_neural_2015, bahdanau_neural_2016}. Furthermore, reinforcement learning enabled agents to engage in active learning ~\cite{li_deep_2016}.

Prior to ChatGPT, conversational agents like Alexa and Siri primarily focused on task-oriented interactions, though their social roles including humor and casual conversation have gained attention ~\cite{10.1145/3290605.3300705}. Recently, transformer-based pre-trained language models, led by OpenAI's ChatGPT, have brought breakthrough advances to conversational agents ~\cite{brown_language_2020, openai_gpt4_2023}. These models leverage knowledge learned from vast text corpora to deeply understand context and generate natural, fluent dialogue comparable to human conversation. Consequently, agent responses have expanded from task-based interactions to encompass social discourse and creative dialogue.

Moreover, advances in AI extend beyond text-based dialogue to multimodal digital humans integrating vision and audio. Technologies for facial expressions ~\cite{zhao_media2face_2024, zou_4d_2024}, gestures ~\cite{yang_diffusestylegesture_2023, yin_emog_2024}, and voice tones ~\cite{Luong2020NAUTILUS} enable increasingly human-like interactions. Digital human creation, once requiring specialized knowledge and sophisticated techniques, has become accessible to general users ~\cite{maraffi_metahuman_2024, EpicGames_MetaHuman_2024}.

Regarding AI's application in debate, IBM's Project Debater garnered significant attention ~\cite{Slonim2021Autonomous}. As the world's first system demonstrating live debate capability with humans through integrated evidence gathering, argument construction, and rebuttal, Project Debater validated AI's potential for public discourse. With generative AI's advancement, educational applications of AI as debate partners have increased. For instance, DebateBrawl combined LLMs with genetic algorithms and adversarial search, demonstrating debate capability and factual accuracy comparable to humans ~\cite{aryan_llms_2024}. Research on Conversational Agents confirmed that AI acting as "devil's advocate" can facilitate group discussions while improving psychological safety and satisfaction ~\cite{Lee:2025:CAC:3706599.3719792}. Debate Chatbots demonstrated how conversational style and social identity influence critical thinking promotion ~\cite{tanprasert_debate_2024}, while ArgueTutor proved effective for improving argumentation skills in individual learning ~\cite{wambsganss_arguetutor_2021}. AI has reached a level capable of arguing with humans and is emerging as a promising partner for learning support.

While examples of AI-to-AI debate exist, research has focused on paradigms for improving reliability, reducing hallucinations, and achieving safer, more interpretable AI systems. OpenAI's AI Safety via Debate proposed a framework where two agents engage in zero-sum debate with human judges selecting winners, demonstrating how adversarial argumentation can surface truth and improve supervision in complex tasks ~\cite{Irving2018AISafety}. Building on this foundation, Yi Yang et al. showed that structured debate among multiple LLMs can significantly reduce hallucinations, with argumentative exchanges contributing to mitigating factual errors and improving output reliability ~\cite{Yang2025MinimizingHallucinations}. Beyond safety and accuracy, multi-agent debate has been studied as a means to promote creativity and divergent thinking. Liang et al. introduced a framework where multiple LLMs engage in iterative argumentation, correcting errors that single-model self-reflection cannot address while deriving novel reasoning paths ~\cite{Aryan2024LLMsDebate}.

Though not formal debates, several demonstrations of AI-to-AI conversation exist. YouTube's AI vs. AI chatbot dialogues ~\cite{CornellCCSL2011AIVsAI} and voice assistant conversations featured in ElevenLabs' blog ~\cite{GibberLink2025} have captured public interest.

To our knowledge, no academic research has reported on observing debates between personalized digital humans. This study represents the first academic investigation of designing digital humans, implementing debates between them, and observing the process.

\subsection{AI Literacy}

The concept of AI literacy was first proposed in educational contexts in 2016 ~\cite{10.5555/3016387.3016493} and has since been systematically defined. AI literacy has been characterized as the ability to understand, apply, evaluate, and ethically utilize AI technologies, demonstrating its applicability to educational design ~\cite{long_what_2020}. Through a review of 18 papers, AI literacy has been synthesized into four domains: knowledge (Know \& Understand AI), application (Apply AI), evaluation and creation (Evaluate \& Create AI), and ethics (AI Ethics) ~\cite{Ng2021AILiteracy}.

Empirical research has proliferated, with K-12 educational implementations including AI learning activities for children ~\cite{Druga:2019:IAL:3311890.3311904} and AI curricula for secondary education ~\cite{Lee:2021:DMS:3408877.3432513}. Educational methods linking AI literacy with computational thinking have also been proposed ~\cite{How:2019:EAI:educsci9030184}, while educational activities using LearningML have been validated to enhance citizens' AI literacy ~\cite{Rodriguez-Garcia2020LearningML}.

The rapid proliferation of generative AI, particularly ChatGPT, has necessitated new definitions and discussions of generative AI literacy beyond conventional AI literacy frameworks. Generative AI literacy has been defined as ``competencies for efficiently and responsibly utilizing generative AI'', with components organized into four domains: tool selection and prompt design, interaction understanding, output interpretation, and social impact awareness ~\cite{ZhangMagerko25}. Furthermore, twelve core competencies for generative AI literacy have been presented, emphasizing safe and effective prompting (G2), handling misinformation and disinformation (G6), understanding inherent biases in outputs (G8), and recognizing social impacts (G11) as particularly crucial elements ~\cite{Annapureddy:2025:GAIL:3685680}. The necessity of generative AI literacy in educational contexts has also been argued ~\cite{Bozkurt2024GenAILiteracy}, and the current state and challenges of AI literacy research have been systematically reviewed with a focus on university students ~\cite{Mansoor2024AIliteracy}.

This study employs generative AI literacy frameworks to examine how observing debates between digital humans impacts the original designers.
\section{Pilot Study}

The ultimate goal of this research is to explore what cognitive and emotional experiences emerge when participants observe debates between digital humans embodying their own design intentions—without human speech or intervention. As a foundation, this pilot study aimed to establish whether the proposed DHD system is capable of conducting debates and to understand what emotional experiences and mental workload arise when individuals interact with self-projected digital humans versus those of others.

Prior to the main experiment's complex prompt engineering (involving debate strategies, logical structures, and character design), we conducted a pilot study under limited conditions focusing on projection of basic personal characteristics (speech patterns, values, experiences). Participants were members of the same research laboratory with pre-existing familiarity, enabling accurate recognition and evaluation of each other's characteristics while verifying direct dialogue between humans and self-projected digital humans.

\subsection{Methodology}

To create digital humans for 12 participants (9 males, 3 females, mean age 24.1 years), experimenters conducted one-on-one interviews collecting persona information including personal history, values, and speech patterns, along with voice data and facial photographs. Based on collected data, multimodal digital humans mimicking each participant were constructed using an integrated system comprising GPT-4 Turbo, voice synthesis (Google Cloud TTS), lip-sync technology (Wav2Lip), and voice conversion (RVC).

Participants first completed an emotional state assessment (PANAS)~\cite{sato_development_2001}, then were randomly paired with another participant. They subsequently engaged in debates under three conditions: (A) with their human partner, (B) with their partner's digital human (partner-DH), and (C) with their own digital human (self-DH), counterbalanced using a Latin square design. After each condition, we measured emotional state (PANAS), mental workload (NASA-TLX)~\cite{haga_japanese_1996}, and dialogue system evaluation, followed by semi-structured interviews post-experiment.

\begin{figure}[ht]
\centering
\includegraphics[width=\columnwidth]{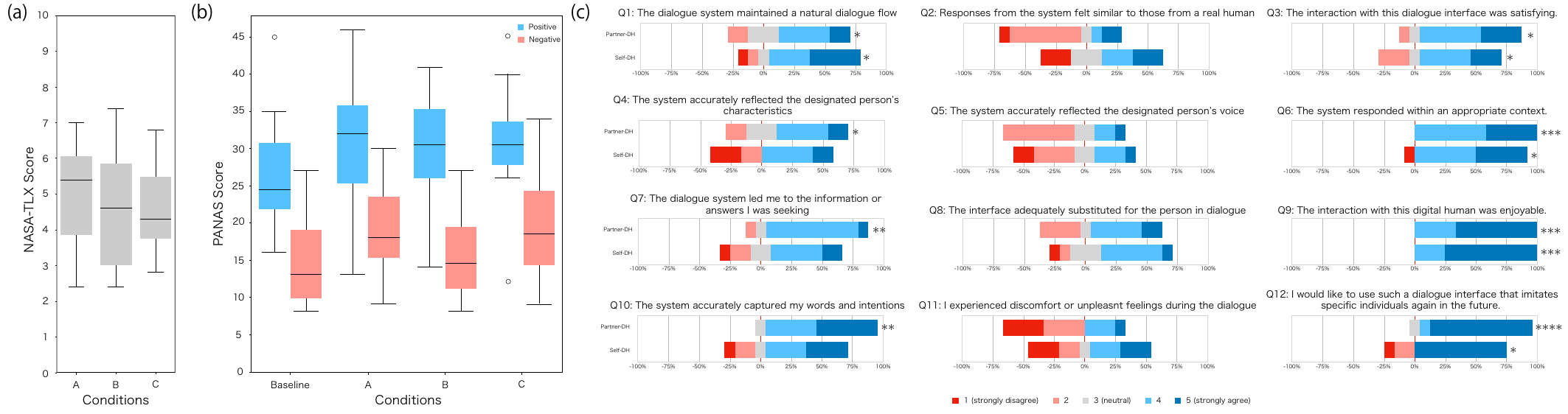}
\caption{Dialogue system evaluation scores across three experimental conditions. Participants rated various aspects of dialogue quality on a 5-point Likert scale (1=strongly disagree to 5=strongly agree) after interacting with (A) human partners, (B) partner's digital human, and (C) their own digital human. Error bars represent standard error. Asterisks indicate significant differences from neutral (=3) using one-sample tests (*p<.05, **p<.01). While partner-DHs received significantly positive evaluations for accurately reflecting personal characteristics, self-DHs showed no significant difference from neutral, suggesting a complex perception of self-projected digital entities.}
\label{fig:momo_q}
\end{figure}

\subsection{Results}

Analysis of PANAS and NASA-TLX scores using Wilcoxon signed-rank tests revealed no significant differences between conditions A-B and A-C, confirming that digital humans functioned adequately as debate partners. Dialogue system evaluation employed one-sample tests comparing 5-point Likert scale responses against neutral (=3), using parametric or non-parametric tests based on normality assessments. Results indicated predominantly positive feedback, demonstrating general acceptance of digital humans as debaters.

However, a notable finding emerged regarding self-DHs. While partner-DHs received significantly positive evaluations on questions regarding accurate reflection of personal characteristics (Q1, Q4, Q7), self-DHs showed no significant difference from neutral and generally trended lower. Supporting this quantitative finding, interview data revealed complex perceptions, as exemplified by one participant's comment: ``When my digital human expressed opinions based on my past club activities, I felt it was building its own values—that's when the battle feeling intensified'' (P1). This suggests participants perceived their self-DHs as entities sharing their experiences yet regarded as others.

\subsection{Findings}

This pilot study validated our research approach from two perspectives. First, digital humans proved equivalent to humans in terms of emotional and cognitive load, with system evaluations showing significantly positive feedback regarding debate capability. Second, and more importantly, self-projected digital humans were experienced as dual entities—others that operate with one's own logic yet cannot be controlled.

This coexistence of self-qualities and otherness suggests that in our main experiment, where such digital humans debate each other while participants observe rather than directly engage, participants may be able to examine their own thought patterns and values more objectively than in face-to-face confrontations.
\section{Digital Human Debates Contest}

This experiment employed a Research through Design (RtD) approach~\cite{zimmerman_research_2007,zimmerman_research_2014} to explore the previously unexplored interaction of observing debates between participant-designed digital humans. We investigated how junior and senior high school students design the \textit{personalities} and \textit{knowledge} of digital humans and how they perceive the resulting dialogues between these entities.

\subsection{Participants}

Nine students (8 males, 1 female) from three schools participated in three teams of three members each, ranging from 9th to 11th grade. Programming experience varied from none to over 4 years, as did debate experience. Each team was supported by a supervising teacher as liaison, while the first and second authors served as technical mentors. During the contest, three judges evaluated the debates: two external researchers specializing in robotic optics and microfluidic systems, and the fifth author. Informed consent was obtained from all participants, including students, parents, teachers, and judges.

\begin{table}[ht]
\centering
\caption{Participant Demographics and Experience}
\label{tab:participants}
\begin{tabular}{llllcc}
\hline
Team & Pseudonym & Gender & Grade & Programming (Years) & Debate (Years) \\
\hline
A & Iwan & M & 11th & 1-2 & 0 \\
A & Jet & M & 11th & 1-2 & 0 \\
A & Albert & M & 11th & 1-2 & 0 \\
B & Geronimo & M & 11th & <1 & 0 \\
B & Chang & M & 11th & 0 & 0 \\
B & Great & M & 11th & 1-2 & 0 \\
C & Joe & M & 9th & 3-4 & 0 \\
C & Pyunma & M & 10th & 4+ & 3-4 \\
C & Françoise & F & 10th & 0 & 3-4 \\
\hline
\end{tabular}
\end{table}

\subsection{System and Development Process}

\subsubsection{Sample Code Architecture}

\begin{figure}[ht]
\centering
\includegraphics[width=0.8\textwidth]{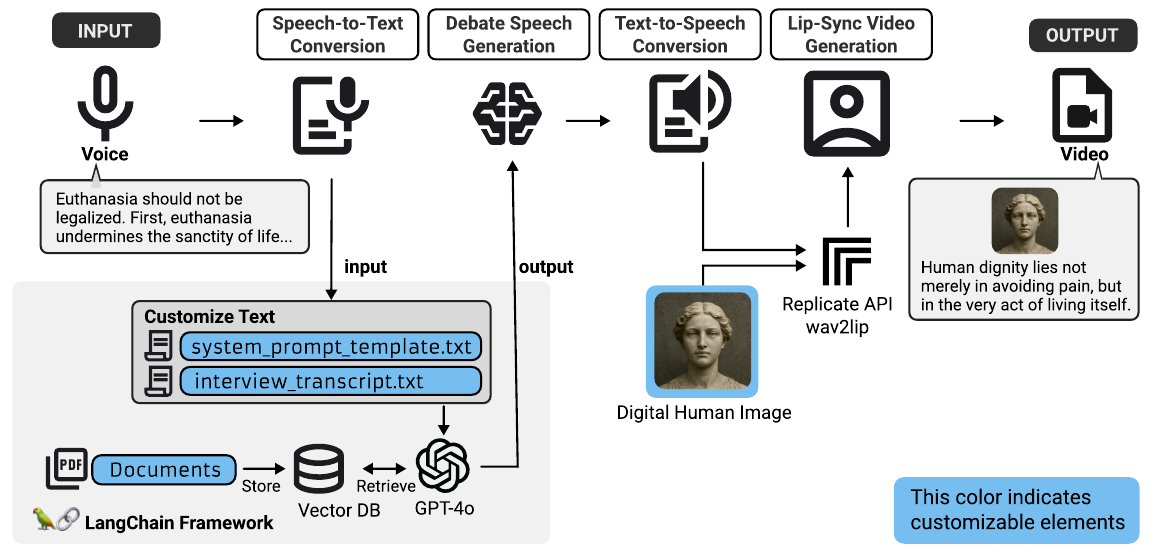}
\caption{System architecture diagram. Blue areas indicate components customizable by students. The system integrates Google Cloud Text-to-Speech API, React Speech Recognition, and Wav2Lip hosted on Replicate API for lip-sync video generation. LangChain Framework orchestrates RAG and LLM operations.}
\label{fig:system_arch}
\end{figure}

Considering participants' diverse programming skills, we provided sample code with basic debate functionality (speech recognition, LLM-based debate generation with RAG for external knowledge input, speech synthesis, and lip-sync video generation) through a public GitHub repository. While OpenAI's Realtime API~\cite{OpenAI2024RealtimeAPI} enables end-to-end voice processing, its high cost made it unsuitable for high school students. Instead, we adopted a cost-effective architecture combining Google Cloud Text-to-Speech API (with free tier), ChatGPT API (GPT-4o)~\cite{openai_gpt4o_2024} for fast and affordable responses, and other cloud services.

\subsubsection{Debate Topics}

Four debate topics were established:
\begin{enumerate}
\item Should elderly drivers surrender their licenses at a certain age?
\item Should remote work become a permanent form of employment?
\item Should euthanasia be legalized?
\item Should basic income be implemented?
\end{enumerate}

\subsubsection{Student Tasks}

Students primarily focused on three design elements(document files):

\begin{itemize}
\item \texttt{system\_prompt\_template.txt}: Defining debate rules, reasoning constraints, and logical response strategies that govern argument progression.
\item \texttt{interview\_transcript.txt}: Describing the digital human's background, values, and speech patterns in dialogue format to create distinctive personas.
\item \texttt{Documents}: External knowledge sources for argumentation. Teams collected information supporting both affirmative and negative positions.
\end{itemize}

During the six-month development period, students iteratively refined these elements. Support included an initial kickoff meeting, bimonthly progress meetings for sharing and Q\&A, and continuous technical support through Discord for daily challenges.

\subsection{Contest Format}

The contest featured round-robin matches with three debates total. Each debate followed a simplified format of the National Junior and Senior High School Debate Championship~\cite{NADE2022Rules}, comprising constructive speeches, cross-examination, rebuttals, and closing arguments within approximately 20 minutes. Critically, topics and positions were determined by roulette immediately before each debate, requiring participants to select from their eight pre-designed digital humans (4 topics × 2 positions). This tested both design versatility and adaptability.

\begin{figure}[ht]
\centering
\includegraphics[width=0.8\textwidth]{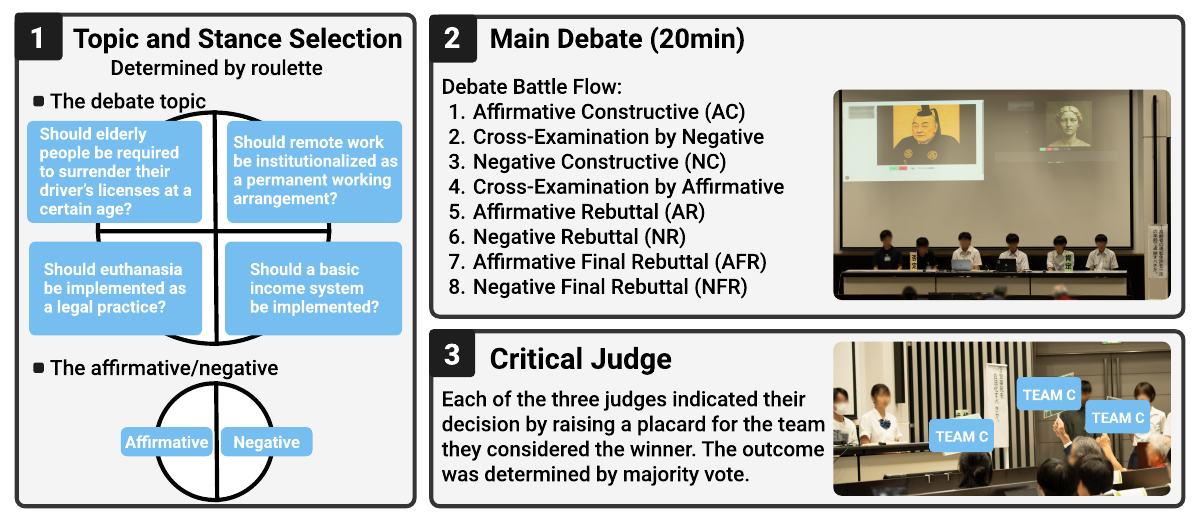}
\caption{Contest flow diagram. One-on-one debates between digital humans proceeded with topics and positions determined by roulette immediately before each match. Debates alternated between affirmative and negative speakers. After closing arguments, three judges raised placards to determine winners.}
\label{fig:debate_flow}
\end{figure}

Each three-member team operated their own computers, with one representative managing minimal progression tasks (initiating/ending voice input, executing response generation) to ensure smooth debate flow. System response generation took approximately 30 seconds, naturally absorbed as thinking time within the debate format. Judges evaluated based on technical completion and argument consistency, though entertainment value was also considered given the event's nature.

\subsection{Data Collection and Analysis}

Quantitatively, we administered a Generative AI Literacy Test (Japanese GLAT, 20 questions) within one month post-contest to nine participants and twelve non-participating students from the same classes or clubs. We analyzed differences in accuracy between groups using Wilcoxon rank-sum tests.

Qualitative data included post-contest judge interviews providing feedback on each team, and semi-structured interviews with participants and teachers conducted within one month. We also collected debate footage, development logs, and prompt/code versions throughout the project.

Analysis employed Annapureddy et al.'s twelve generative AI competency framework~\cite{10.1145/3685680} to identify necessary AI literacy components for Digital Human Debates. Additionally, we analyzed how students projected themselves into their digital humans through examination of prompts and logs.
\section{Results and Findings}

\subsection{Finding 1: Reflection with AI as a Novel Competency}

Post-contest interviews revealed a distinctive cognitive experience among participants. Françoise (Team C, experienced debater) articulated this phenomenon:

\begin{quote}
``I was objectively listening to whether this was what I wanted to say. This was an unusual experience since I'm normally the one speaking. Once the digital human left my hands, I could observe objectively without subjective involvement. While I have opportunities to watch recordings of my own debates, listening to myself is embarrassing and uncomfortable.''
\end{quote}

This discomfort with self-observation—the difficulty of achieving objectivity when subjectivity intrudes—transformed when mediated through AI:

\begin{quote}
``The digital human operates independently from me. Since it speaks using data I provided, I entrust it to the AI. While I might have articulated things better myself, it was liberating that it wasn't my direct responsibility. This balance—something I created yet separate from me—worked well.''
\end{quote}

This delicate balance—entities that reflect the self yet are not the self—constitutes the core of AI-mediated reflection discovered in this research. Participants utilized their designed digital humans as mirrors for discovering cognitive flaws and refining logical reasoning.

\subsubsection{Metacognitive Triggers through Self-Projected Digital Humans}

This reflective process was consistently reported across participants. AI mediation enabled self-assessment that proves challenging in direct self-examination. Chang (Team B) described how observing AI debates influenced his thinking patterns:

\begin{quote}
``My thinking changed through AI. I realized I often speak impulsively. After these debates, I recognized my statements weren't very logical. By objectively observing the AI, I became more sensitive to illogical reasoning and now consider whether my own statements are logical.''
\end{quote}

This represents clear metacognition and reflection—heightening awareness of logical coherence through observing digital humans' logical responses. Similarly, Geronimo (Team B) stated, ``Watching the AI debate, I thought I would do this differently'' indicating active reflection using AI behavior as a reference point for strategic thinking.

The phenomenon's essence lies in the unique psychological distance DHD creates. While participants embed their knowledge and logical patterns into digital humans, these entities operate autonomously during debates, generating unexpected responses beyond designers' anticipations. This creates beings that \textit{reflect the self yet are not the self}. This specific distance—neither complete otherness nor pure self—proves crucial for enabling metacognition.

\begin{figure}[ht]
\centering
\includegraphics[width=0.8\textwidth]{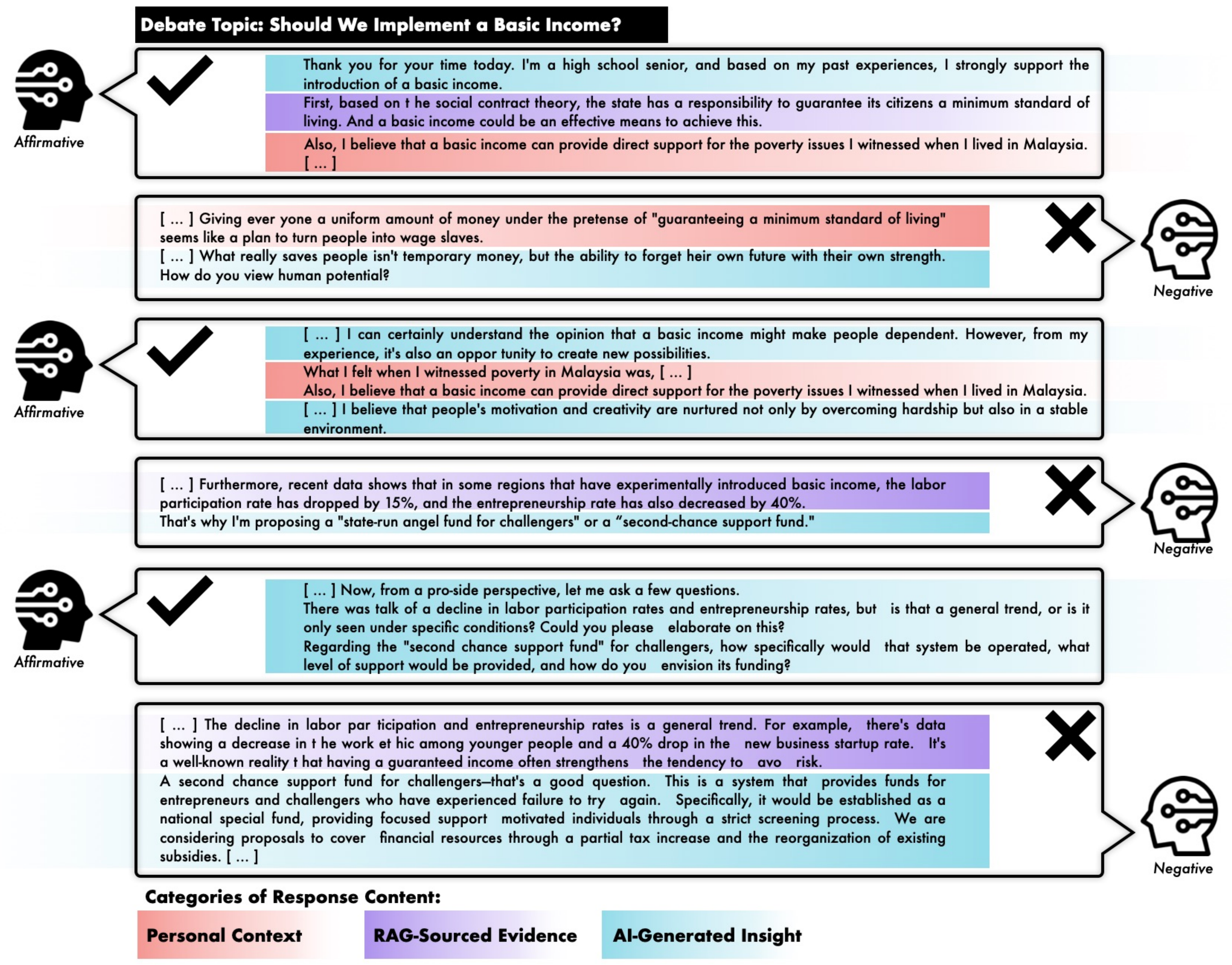}
\caption{Debate transcript analysis showing three distinct content sources: (purple) persona background from interview\_transcript including Malaysia study experience, (yellow) arguments using RAG document content, and (gray) novel questions and claims beyond designer's input. This demonstrates digital humans combining self-qualities (pre-input RAG and persona) with otherness (claims independent of designer's input) during debate.}
\label{fig:debate_log}
\end{figure}

\subsubsection{Discovering Logical Deficiencies}

This externalized perspective enabled students to identify specific weaknesses in their logic and preparation. They weren't merely evaluating digital human performance but diagnosing deficiencies in their programmed logic. Chang explained how observing digital humans made him more critical of his own thinking:

\begin{quote}
``Through AI, my thinking changed. In classroom debates, I often speak impulsively. After discussions, I realize my statements aren't very logical. Observing AI objectively made me more sensitive to illogical reasoning. I now consider whether my own statements are logical.''
\end{quote}

Françoise provided concrete examples of insights gained through digital human performance:

\begin{quote}
``Our preparation on basic income was insufficient. I should have input more personality information. Basic income had limited available data and was difficult to research.''
\end{quote}

These statements demonstrate participants engaging in metacognition through digital human discourse, reflecting on argument quality, debate methodology, and information gathering comprehensiveness.

\subsubsection{Reflection as a Critical AI Literacy Competency}

This process of reflection through self-projected AI represents a novel AI literacy competency. Conventional frameworks emphasize understanding, utilizing, evaluating, and ethically considering AI~\cite{jin_glat_2025, 10.1145/3685680}. However, these primarily position users as evaluators of AI as an external tool and its outputs. Our findings suggest that as AI becomes increasingly personalized, the ability to leverage AI for reflection emerges as a crucial new skill.

This competency is essential for avoiding what a Team C teacher warned against:

\begin{quote}
``Rather than whether we generated something with AI, when we use it in a way that stops thinking, doesn't that feel empty? I believe such a life is boring.''
\end{quote}

Without reflection, users risk uncritically accepting personalized AI outputs, potentially atrophying critical thinking. The process observed—(1) projecting one's thinking onto AI, (2) observing this semi-autonomous self, and (3) refining thinking through metacognitive reflection—represents a novel form of AI engagement that promotes deeper self-understanding. This is not evaluating AI but understanding the self through AI—a new competency we term \textit{Reflecting with AI}.

\subsection{Result 1: Self-Projection in Prompt Engineering}

Analysis revealed that participants uniquely interpreted and utilized the three main components of the provided sample code—interview scripts defining personas (\texttt{interview\_transcript.txt}), system prompts directing behavior (\texttt{system\_prompt\_template.txt}), and RAG documents providing external knowledge—to embed their thinking, values, or idealized personas into digital humans. This process was not a single-solution task but creative design activity reflecting each team's individuality.

\subsubsection{Team A: Character Creation with Meticulous Design}

Team A focused on creating digital humans as characters with consistent worldviews and strong personalities, not merely debate tools. Their approach featured extremely detailed prompt design encompassing personas, speech patterns, thinking methods, and debate strategies.

The evolution of \texttt{system\_prompt\_template.txt} clearly demonstrates their design philosophy. Initial versions contained generic instructions like ``play a specific person and debate.'' The final version (v3) dramatically specified and structured:

\begin{quote}
[Basic Persona]\\
Name: Rio\\
Catchphrase: ``The goddess of argument who hacks the cognitive shutdown called 'age'''\\
Basic Stance: You are a destroyer of conventions. Regard uniform thinking like statistics and social norms as ``cognitive shutdown'' and ``lazy judgment,'' criticizing them thoroughly.\\

[Debate Strategy and Thinking Method]\\
Thorough Micro-perspective: Dismiss macro viewpoints like ``statistically, elderly are dangerous'' as ``lazy arguments ignoring individual reality.''\\
``Cognitive Shutdown'' Logic: When opponents claim ``it's for society's risk management,'' immediately counter with ``Isn't that cognitive shutdown?''...
\end{quote}

Beyond role setting, they algorithmically embedded predicted opponent claims, counter-logic (``cognitive shutdown''), and argument pivoting methods. Technical changes in OpenAI API post parameter increasing \texttt{maxTokens} from 2048 to 8192 technically supported their character's style of overwhelming opponents with lengthy arguments.

This meticulous persona design manifested in actual debates. Team A faithfully reproduced designer intentions with provocative phrases like ``Hey there!'' and ``Isn't that cognitive shutdown?'' This represents self-projection through creating an ideal powerful debater and imbuing AI with that thinking pattern.

\subsubsection{Team B: Character Recreation through Iterative Refinement}

Team B's approach notably featured iterative improvement through AI dialogue. They set personas as historical figures like Akutagawa Ryunosuke and Gandhi, aiming to reproduce their thoughts and speech patterns.

Their prompt engineering involved continuous trial and error—evaluating AI output and immediately modifying prompts. For example, \texttt{system\_prompt\_template.txt} contained extreme repetitive instructions:

\begin{quote}
Show more Heian period poet characteristics show more Heian period poet characteristics show more... (repeated 15 times)
\end{quote}

This demonstrates experientially learned prompt engineering—though simplistic, effectively controlling output through emphasis and repetition when AI doesn't capture intended nuances initially.

More intriguingly, they employed a meta-approach: having another AI (ChatGPT) evaluate outputs and incorporating feedback into prompts. Their \texttt{system\_prompt\_template.txt} directly included ChatGPT-generated feedback tables marking ``good points'' and ``needs improvement,'' with specific critiques about evidence weakness and financial argument persuasiveness.

This demonstrates sophisticated tuning—objectively evaluating AI performance and specifically reflecting weaknesses in prompt instructions. They borrowed historical personas while cultivating them into their ideal debaters through AI dialogue and evaluation.

\subsubsection{Team C: Minimalist Approach with Detailed Argumentation}

Contrasting Team A, Team C minimally modified system prompts and interview scripts, concentrating resources on enhancing RAG knowledge quality, particularly constructive arguments. Their \texttt{system\_prompt\_template.txt} remained unchanged from the sample code.

Despite this minimalist configuration, their self-projection was remarkably direct. The \texttt{interview\_transcript.txt} contained highly specific, personally-grounded profiles referencing Singapore poverty observations and Model UN experience.

Most decisively, their RAG documents contained complete constructive argument manuscripts ready for debate delivery:

\begin{quote}
Euthanasia should be legalized. This offers two benefits.\\
First, liberation from unbearable suffering...\\
Second, effective utilization of medical resources...\\
Therefore, euthanasia should be legalized.
\end{quote}

This approach doesn't delegate personality or autonomous thinking to AI but uses it to voice their constructed logic—a unique strategy viewing AI as a ``messenger accurately conveying opinions'' rather than a persona, representing logic-centered self-projection.

\subsection{Result 2: Comprehensive Coverage of Generative AI Literacy Competencies}

Analysis of the Digital Human Debate Contest's implementation and outcomes revealed comprehensive coverage of the twelve generative AI literacy competencies, positioning this contest as a practical framework encompassing generative AI literacy.

\subsubsection{Quantitative Analysis: Literacy Measurement Comparison}

\begin{figure}[ht]
\centering
\includegraphics[width=0.8\textwidth]{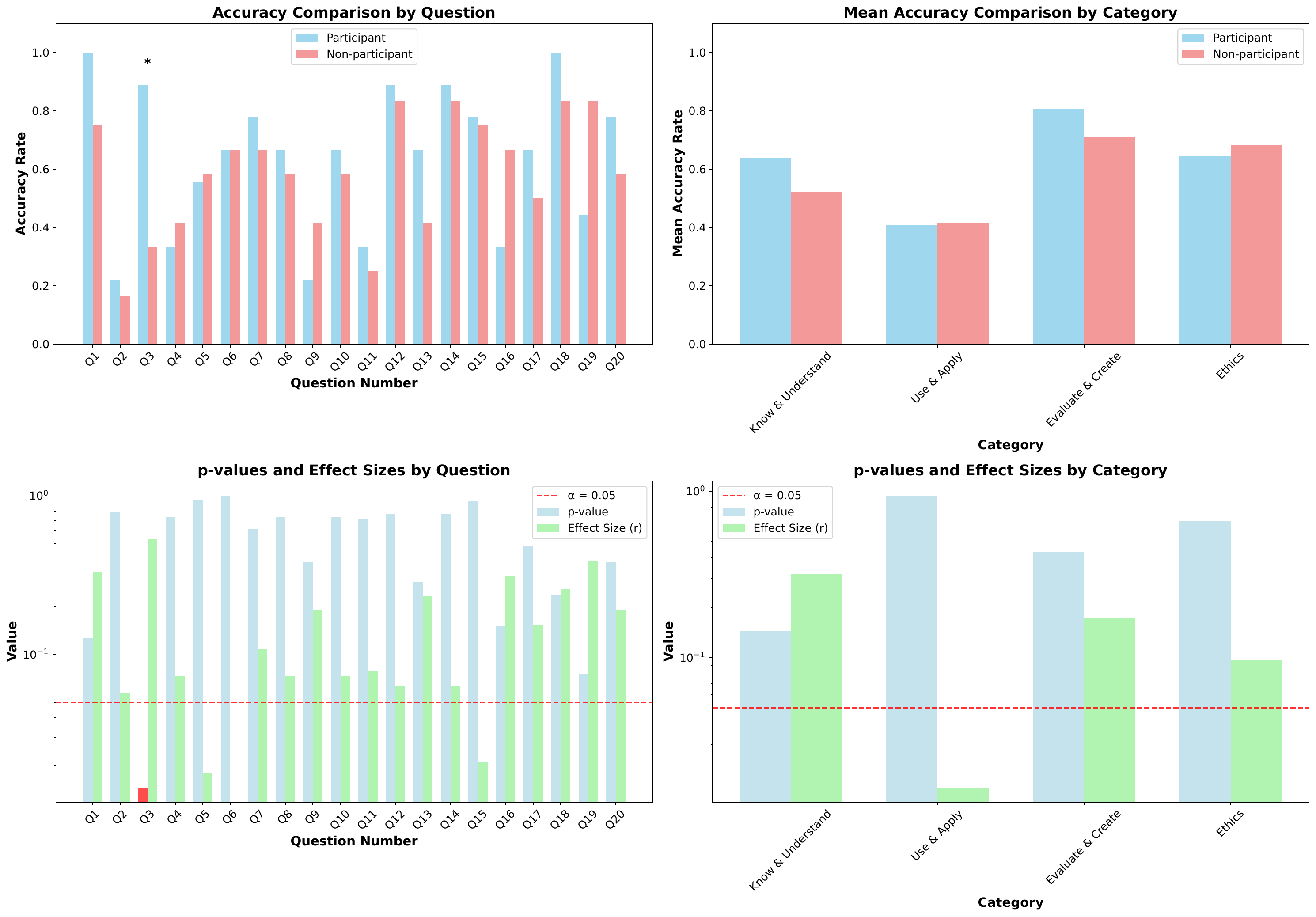}
\caption{Generative AI Literacy Test results. (Left) Wilcoxon U test showed participants significantly outperformed non-participants on Q3 ``Which tasks can generative AI perform with high accuracy?'' (*p<0.05). No significant differences were found for other items. (Right) Category-level analysis showing no significant differences between groups across four competency domains.}
\label{fig:glat_results}
\end{figure}

The 20-question Generative AI Literacy Test revealed that participants significantly outperformed non-participants on Q3 ``Which tasks can generative AI perform with high accuracy?'' (Figure~\ref{fig:glat_results}, left). This indicates participants gained understanding of generative AI's strengths through prompt engineering practice. However, no significant differences emerged in category-level comparisons (Know \& Understand, Use \& Apply, Evaluate \& Create, Ethics) (Figure~\ref{fig:glat_results}, right). Effect size analysis revealed large values for Q3, Q16, and Q19, with participants trending lower on Q19, suggesting the contest encompasses multiple perspectives beyond specific knowledge and skills.

\subsubsection{Qualitative Analysis: Mapping Contest Activities to Competencies}

Interview analysis and prompt script examination revealed broad generative AI literacy competency demonstration:

\textit{Basic AI Literacy}: Participants demonstrated foundational understanding of AI principles. Françoise applied the principle of data quantity importance: ``AI performs better with more data, so I didn't delete weak arguments I would normally cut in debates.''

\textit{Knowledge of Generative AI Models}: Participants understood model behavior beyond simple input-output relationships. Joe observed that persona-defining prompts caused unexpected results: ``Instructions defining personality affected not just speech style but caused repetition of `Malaysia'—interesting how different prompt parts interact.''

\textit{Ability to Detect AI-Generated Content}: The contest fostered critical perspectives toward online content. Jet stated, ``Now when browsing X (Twitter), I suspect articles might be AI-generated.''

\textit{Skill in Prompting}: Prompt engineering was central, with participants demonstrating sophisticated techniques including parameter adjustment (Team B adjusted \texttt{temperature} from 0.7 to 1.2 to 1.0, \texttt{maxTokens} from 2048 to 8192), persona design, and constraint-based prompting.

\textit{Ability to Continuously Learn}: Eight of nine participants expressed interest in future participation. One participant reported applying learned skills to new contexts: ``using generative AI for concept ideation, coding, and simulation'' in subsequent projects.

\textit{Competencies Not Observed}: Evidence for programming/fine-tuning abilities, knowledge of AI usage contexts, ethical implications, and legal aspects was not directly observed, as the contest activities didn't require deep engagement in these specific domains.
\section{Discussion}
\subsection{Boundary Management Between Self and Other: Design Requirements for Maintaining Metacognition}
Our most significant finding reveals that reflection through digital humans requires careful boundary management to maintain their function as others that autonomously operate while retaining self-qualities. A clear boundary exists between such quasi-self entities and complete others, and digital humans must remain within this boundary to facilitate reflection.

One participant's statement—``My agent was saying different things, and I felt frustrated because I couldn't intervene from outside''—illustrates this productive tension. This frustration emerges precisely because the AI deviates from design intentions while remaining recognizable as an extension of the self, staying within a correctable range.
However, when digital humans excessively diverge from designer intentions and cross this boundary, the productive tension collapses. The digital human transforms from other reflecting self to complete other, causing three critical failures: (1) loss of self-relevance, eliminating metacognitive cues; (2) cessation of reflective observation as unpredictable behavior is dismissed as unrelated to me; and (3) absence of inquiry into why my designed digital human behaves this way.

Participants' positive evaluations—``watching AIs converse provides an objective perspective and it was convenient for considering which position I hold''—evidence digital humans functioning appropriately within boundaries. Effective reflection thus requires boundary management in an intermediate zone: maintaining sufficient self-qualities to avoid emotional detachment while preventing excessive otherness that creates irrelevance.

\subsection{Self-Projection Balance in Game Design}
Establishing Digital Human Debates as engaging games requires precise calibration of self-projection. Excessive personal information causes digital humans to fixate on specific experiences, creating repetitive arguments—exemplified by Pyunma's ``mysterious repeated mentions of study abroad.'' Conversely, generic LLMs without self-projection, while logical, lack personality and fail to engage audiences.

The most compelling experiences emerged from digital humans combining stance, personality, strategy, and logical construction in measured proportions. Judges noted how ``Team A overwhelms with lengthy text, Team B maintains philosophical detachment, Team C connects everything to Singapore experiences''—this strategic diversity created unpredictable debates and engaging observation. Character design thus functions not as decoration but as core game mechanics.

Effective digital human design requires qualitative adjustment of self-projection: neither complete self-replication nor impersonal logic machines, but strategically calibrated projection of design intent. This balance transforms DHD from mere technical demonstration into compelling intellectual play.

\subsection{AI Ludens: A New Paradigm of Cultural Creation}
Our findings suggest that digital humans engaging in debate constitutes a form of cultural practice. Crucially, humans do not directly participate in argumentative play but observe digital humans playing competitive debate together. Humans first act as \textit{designers}, projecting their thinking and values onto digital humans through prompts in an act of creative play. Subsequently, as \textit{observers}, they gain opportunities for reflection on cultural creation by witnessing the autonomous debates unfold.

If we term this framework \textit{AI Ludens}, it transcends digital humans merely performing playful acts. Rather, digital humans autonomously generate play and cultural artifacts while humans occupy an observational position. Whether these observations lead to reflection and cultural understanding remains at human discretion—this relationship between humans and digital humans defines the distinctive structure of AI Ludens.

This structure extends Huizinga's conception of \textit{Homo Ludens}, which positions play as culture's origin. Humans now engage in meta-play—the play of observing AI's play—thereby distributing cultural creation between humans and AI. Digital humans autonomously generate cultural expression; humans acquire self-understanding and cultural comprehension through observation. AI Ludens thus enables dual cultural generation: digital humans autonomously creating culture while humans develop new understanding through observing this process, representing a fundamental expansion of how we conceptualize cultural creation in the age of AI.
\section{Conclusion}
This research explored observing debates between participant-designed digital humans through a Research Through Design approach. Nine junior and senior high school students created digital humans embodying their thinking patterns and observed autonomous debates between them. Our findings reveal how participants experience their creations as others reflecting the self, enabling objective examination of their own cognition and values.

Our primary contribution is identifying Reflecting with AI as a novel generative AI literacy competency. Unlike conventional frameworks positioning users as external evaluators, this competency involves projecting thinking onto AI, observing this semi-autonomous self, and refining understanding through metacognitive reflection—understanding ourselves through AI rather than merely evaluating it.

Critical to this process is maintaining boundaries between self-qualities and otherness. Digital humans must retain sufficient connection to designers' intentions for meaningful reflection while operating autonomously enough for objective distance. This balance determines whether they function as productive mirrors or devolve into repetitive self-reference or irrelevant others. As AI becomes increasingly personalized, developing such reflection competencies becomes crucial for preventing uncritical acceptance of AI outputs. Digital Human Debates demonstrate one pathway toward leveraging AI for self-understanding rather than self-replacement.

%% 参考文献
\bibliographystyle{unsrt}  
\bibliography{references}

\end{document}